\documentclass[twocolumn,aps,prd,showpacs,preprintnumbers,floatfix,nofootinbib,superscriptaddress]{revtex4}
\usepackage{graphicx}
\usepackage[]{latexsym,amssymb,amsmath}
\newcommand{\be}{\begin{eqnarray}}
\newcommand{\ee}{\end{eqnarray}}
\renewcommand{\d}{\mbox{${\rm d}$}}
\newcommand{\ein}{\epsilon_{-}}
\newcommand{\eout}{\epsilon_{+}}
\newcommand{\einz}{{\epsilon_0}_{-}}
\newcommand{\eoutz}{{\epsilon_0}_{+}}

\newcommand{\ainz}{{a_0}_{-}}
\newcommand{\aoutz}{{a_0}_{+}}

\newcommand{\h}{\mathcal{H}}
\newcommand{\ezr}{\epsilon_0^{\rm rad}}
\newcommand{\ezd}{\epsilon_0^{\rm dust}}
\newcommand{\lp}{\ell_{\rm P}}
\newcommand{\mpl}{M_{\rm P}}
\begin{document}
\title{Bubble dynamics: (nucleating) radiation inside dust}
\author{R.~Casadio}
\email{casadio@bo.infn.it}
\affiliation{Dipartimento di Fisica, Universit\`a di Bologna, via Irnerio 46, 40126 Bologna, Italy}
\affiliation{INFN, Sezione di Bologna, Via Irnerio 46, I-40126 Bologna, Italy}
\author{A.~Orlandi}
\email{orlandi@bo.infn.it}
\affiliation{Dipartimento di Fisica, Universit\`a di Bologna, via Irnerio 46, 40126 Bologna, Italy}
\affiliation{INFN, Sezione di Bologna, Via Irnerio 46, I-40126 Bologna, Italy}
\begin{abstract}
We consider two spatially flat Friedmann-Robertson-Walker spacetimes divided
by a time-like thin shell in the nontrivial case in which the inner region of finite extension
contains radiation and the outer region is filled with dust.
We will then show that, while the evolution is determined by a large set of
constraints, an analytical description for the evolution of the bubble radius
can be obtained by formally expanding for short times after the shell
attains its minimum size.
In particular, we will find that a bubble of radiation, starting out with vanishing
expansion speed, can be matched with an expanding dust exterior, but not
with a collapsing dust exterior, regardless of the dust energy density.
The former case can then be used to describe the nucleation of a
bubble of radiation inside an expanding dust cloud, although the final
configuration contains more energy than the initial dust,
and the reverse process, with collapsing radiation transforming into
collapsing dust, is therefore energetically favored.
We however speculate a (small) decaying vacuum energy or cosmological
constant inside dust could still trigger nucleation.
Finally, our perturbative (yet analytical) approach can be easily adapted to different
combinations of matter inside and outside the shell, as well as to more general
surface density, of relevance for cosmology and studies of defect formation
during phase transitions.
\end{abstract}
\pacs{04.40.-b,04.20.-q,04.20.Cv,98.80.Jk}
\maketitle
\section{Introduction}
\label{intro}
The dynamics of an infinitely thin, spherically symmetric shell $\Sigma$
separating two spacetime regions $\Omega_\pm$ with given metrics is a well
known problem of General Relativity.
The general theory dates back to 1965~\cite{israel} and is completely understood.
Given the symmetry of the system, we can use spherical coordinates
$x^\mu_\pm=\{t_\pm,r_\pm,\theta,\phi\}$ in $\Omega_\pm$, respectively,
where the angular coordinates are the same in both patches,
and $0\le r_-<r_-^{\rm s}$, $r_+^{\rm s}< r_+$,
with $r_\pm^{\rm s}=r_\pm^{\rm s}(t_\pm)$ the (in general time-dependent)
radial coordinates of the shell in $\Omega_\pm$.
One then takes specific solutions $g_{\mu\nu}^\pm$ of the Einstein equations
inside $\Omega_\pm$ and imposes suitable junction conditions across $\Sigma$,
namely the metric is required to be continuous across the shell,
\be
\left.g_{\mu\nu}^+\right|_\Sigma
=
\left.g_{\mu\nu}^-\right|_\Sigma
\ ,
\label{cont}
\ee
whereas the extrinsic curvature $K_{ij}$ of $\Sigma$ is allowed to have a jump
proportional to the surface stress-energy tensor of the time-like shell $\sigma_{ij}$
(italic indices run on the shell's three-dimensional world-sheet),
\be
[ K_i^{\ j}]
-
\delta_i^{\ j}\,[ K_l^{\ l}]
=
\kappa\, \sigma_i^{\ j}
\ ,
\label{K}
\ee
in which $[K_i^{\ j}] \equiv K_i^{\ j}|_+ - K_i^{\ j}|_-$ denotes the difference between
extrinsic curvatures on the two sides of the shell.
Note that we set $c=1$ and $\kappa=8\,\pi\,G_{\rm N}/3=\lp/\mpl$ ($=1$ when convenient),
where $G_{\rm N}$ is Netwon's constant and $\lp$ ($\mpl$) the Planck length
(mass).
Although the classical evolution equation~\eqref{K} may appear simple,
it has been solved only in a few special cases, most notably for the vacuum
or with cosmological constants in $\Omega_\pm$~\cite{blau} 
(for an extensive bibliography see Ref.~\cite{ansoldi}).
\par
A most intriguing result emerges in the semiclassical picture~\cite{coleman,farhi},
where one finds that ``bubbles'' can be quantum mechanically created from nothing
(in a sense, at the expense of gravitational energy).
This may occur when one has a classical solution for an expanding shell areal radius
with a (finite) minimum value (turning point of the classical trajectory, larger than $\lp$),
and a non-vanishing quantum mechanical amplitude for the ``tunneling'' into such
a system from one without the shell (that is, a shell of zero area).  
It has been conjectured that these bubbles could represent child universes generated
inside a parent (or ``landscape'') universe~\cite{yun,guth,linde}, if they expand indefinitely
(or at least long enough).
Bubble dynamics might also be used to model regions of space within which a matter
phase transition occurs (from false to true vacuum, as well as between different form
 of matter~\cite{bkt}).
One can, for instance, use such a model to approximate the formation of radiation from a
decaying scalar field during reheating after inflation.
It is known that for an inflaton with a quadratic potential, the time averaged dynamics
of the final oscillation phase mimics that of matter.
The approach developed in this paper could turn out to be suitable to describe the
decay into radiation.
In the end, knowing the correct evolution of such a bubble would be of great help in
understanding how defects formed during phase transitions are ``ironed out'' by the
expansion of the new phase.
\par   
In this paper, we are mainly interested in presenting an analytical (perturbative in time)
approach to study a time-like shell's dynamics and to obtain analytical conditions for
the existence of expanding bubbles in terms of the energy densities inside and outside
the shell, when such regions contain homogeneous dust or radiation.
This problem is made technically cumbersome because of the occurrence
of algebraic constraints (to ensure the arguments of proliferating square roots are positive).
We shall therefore consider in details only the specific case of nucleation of a
spatially flat radiation bubble inside (spatially flat) collapsing or expanding dust,
in order to keep the presentation of our method more streamlined.
Nonetheless, these cases are also of particular physical interest.
For example, one can conceive the density inside a collapsing astrophysical
object might be large enough to allow for the creation of supersymmetric
matter which, in turn, would then annihilate regular matter
and produce a ball of radiation~\cite{clavelli}.
Likewise, in a dense matter-dominated expanding universe, one might consider
the possibility of spontaneous nucleation of radiation bubbles.
We shall find the dust energy density just sets the overall scale of the problem.
Assuming the bubble surface density is (initially) constant,
expanding radiation bubbles may then be matched with an expanding dust exterior,
the time-like shell surface density being uniquely related to the inner radiation density.
In order to view the bubble creation as a phase transition from dust to
radiation (plus the surface density of the shell), one however needs an
external source of energy, since the total energy of the bubble is larger than
the initial energy of the dust.
This implies that the reverse process of collapsing radiation turning into
collapsing dust is actually favored energetically.
Moreover, no configuration with expanding bubbles is allowed inside collapsing dust,
regardless of how large is the dust energy density, and the conversion between the
two types of matter therefore appears highly disfavored in this case.
\par
In Section~\ref{bub_dyn}, we briefly review the fundamental equations and
constraints that describe general time-like bubble dynamics, following Ref.~\cite{bkt},
and then specify all expressions for $\Omega_\pm$ given by spatially flat
Friedmann-Robertson-Walker (FRW) regions filled with dust or radiation.
In Section~\ref{rad_dust}, we work out the explicit case of a radiation bubble of
constant surface density nucleated inside collapsing or expanding dust, for which
we obtain the initial minimum radius in terms of the inner and outer energy densities.
Finally, in Section~\ref{conc}, we make some considerations about our findings
and possible future generalizations.
\section{Bubble dynamics}
\label{bub_dyn}
For our analysis, we will mostly follow the notation of Ref.~\cite{bkt}, where
the metric in each portion $\Omega_\pm$ of spacetime is given by
\be
\d s^2 = e^{\nu(r,t)}\,\d t^2 - e^{\lambda(r,t)}\,\d r^2 - R^2(r,t)\, \d \Omega^2
\ ,
\label{met0}
\ee
where $t=t_\pm$ are the time coordinates inside the corresponding patches,
and likewise for the three spatial coordinates.
On the shell time-like surface $\Sigma$, one has the line element
\be
\left.\d s^2 \right|_\Sigma = \d \tau^2 - \rho^2 (\tau)\, \d \Omega^2
\ ,
\ee
in which $\tau$ is the proper time as measured by an observer at rest with the shell
of areal radius $\rho=R_\pm(r_\pm^{\rm s}(t_\pm),t_\pm)$.
The relation between $\tau$ and $t_\pm$ is obtained from the equation of continuity
of the metric, Eq.~\eqref{cont}, and is displayed below in Eq.~\eqref{tt} for the cases of interest.
On solving Eq.~\eqref{K} in terms of $\rho$ and $\dot\rho=\d\rho/\d\tau$,
one gets the dynamical equation
\be
\dot\rho^2(\tau)=B^2(\tau)\,\rho^2(\tau)-1 
\ ,
\label{evolution}
\ee
where
\be
B^2=\frac{\left(\eout+\ein+9\,\kappa\,\sigma^2/4\right)^2-4\,\ein\, \eout}{9\,\sigma^2}
\label{B}
\ ,
\ee
with $\sigma=\sigma_{\ 0}^0(\tau)$ the shell's surface density and
$\epsilon_\pm=\epsilon_\pm(t_\pm)$ the time-dependent energy densities in
$\Omega_\pm$ respectively.
It is important to recall that metric junctions can involve different topologies
for $\Omega_\pm$, but we are here considering only the so-called ``black hole''
type, in which both portions of spacetime have increasing area radii $R_\pm$
in the outward direction (of increasing $r_\pm$).
Assuming the surface density of the shell is positive, one must then
have~\footnote{This also implies that $\eoutz > \einz$ and,
from the Friedmann equation~(\ref{friedmann1}) given
below, $H_{+}^2 > H_{-}^2$.}
\be
\eout(\tau)-\ein(\tau) > \frac{9}{4}\, \kappa\, \sigma^2(\tau)
\ ,
\label{difference}
\ee
at all times, in order to preserve the chosen spacetime
topology~\cite{bkt, sm}.
\par
In the pure vacuum case, $\epsilon_\pm$ are constant and for constant $\sigma$
the solution is straightforwardly given by
\be
\rho(\tau) = {B^{-1}}\, \cosh (B\,\tau)
\ ,
\label{vacuum}
\ee
where $B=B(\epsilon_\pm,\sigma)$ from Eq.~\eqref{B} is also constant~\cite{blau}.
\par
In the non-vacuum cases, finding a solution is however significantly more involved.
Regardless of the matter content of $\Omega_\pm$, it is nonetheless possible to
derive a few general results for a bubble which nucleates at a time $\tau=\tau_0$,
that is a shell that expands from rest, 
\be
\dot\rho_0=0
\ ,
\label{drho0}
\ee
with initial finite (turning) radius ($\rho_0> 0$), where the subscript $0$ will always indicate
quantities evaluated at the time $\tau=\tau_0$.
First of all, from Eq.~\eqref{evolution}, the initial radius must be given by
\be
\rho_0=\left| B_0^{-1}\right|
\ ,
\label{rho0}
\ee
which requires $B_0$ real, or
\be
\left(\eoutz+\einz+9\,\kappa\,\sigma_0^2/4\right)^2>4\,\einz\, \eoutz
\ .
\label{B0}
\ee
This condition is always satisfied if $\ezd$ and $\ezr$ are both positive and will
therefore be of no relevance in this paper, but must be carefully considered
when allowing for negative energy densities (and non-vanishing spatial curvature).
Further, upon deriving Eq.~\eqref{evolution} with respect to $\tau$
(always denoted by a dot) 
\be
2\,\dot\rho\,\ddot\rho=2\left(B\,\dot B\,\rho^2+B^2\,\rho\,\dot\rho\right)
\ ,
\ee
and using Eq.~\eqref{drho0}, one also obtains
\be
\dot B_0=0
\ ,
\label{dB0}
\ee
assuming $\ddot\rho_0$ is not singular.
The constraint in Eq.~\eqref{difference} at $\tau=\tau_0$,
\be
\eoutz-\einz > \frac{9}{4}\, \kappa\, \sigma_0^2
\ ,
\label{difference0}
\ee
and the conditions in Eqs.~\eqref{drho0} and \eqref{dB0}
will play a crucial role in the following.
\subsection{Flat FRW regions}
Since we wish to study the particular case of a shell $\Sigma$ separating two 
regions  $\Omega_\pm$ filled with homogeneous fluids, the metrics in
$\Omega_\pm$ will be taken to be the usual FRW expressions.
Moreover, we already assumed $\Omega_-$ has finite initial
extension and, by definition, represents the interior of the shell.
As a further simplification, we shall only consider flat spatial
curvature and set the cosmological constant $\Lambda = 0$
everywhere.
\par
The metrics~\eqref{met0} in the inner and outer regions are therefore given by
\be
\d s^2
=
\d t^2
-{a^2}{(t)}\left[
\d r^2
+ r^2\left(\d\theta^2+\sin^2\theta\,\d\phi^2\right)
\right]
\ ,
\label{frwmetric}
\ee
where $a(t)$ is the scale factor which evolves according to
the Friedmann equations 
\be
&
\strut\displaystyle
H^2
=
\left(\frac{1}{a}\,\frac{\d a}{\d t}\right)^2
=
\kappa\, \epsilon
&
\label{friedmann1}
\\
&
\strut\displaystyle
2\,
\frac{1}{a}\,\frac{\d^2 a}{\d t^2}
+
\left(\frac{1}{a}\,\frac{\d a}{\d t}\right)^2
=
-3\,\kappa\, p
\ .
&
\label{friedmann2}
\ee
We assume the energy density $\epsilon$ and pressure $p$ 
of the fluids obey barotropic equations of state,
\be
p=w\,\epsilon(t)
\ ,
\ee
and recover the well-known behaviors
\be
\epsilon(t)\, \left(\frac{a(t)}{a_0}\right)^{3\,(w+1)}=\epsilon_0
\label{conservation}
\ ,
\ee
in which $\epsilon_0$ is the density evaluated at a reference instant of time
$t=t_0$ and $a_0=a(t_0)$.
For dust, $w=0$ ($p=0$), whereas for radiation $w=1/3$, so that
\be
\begin{array}{l}
\epsilon^{\rm dust}(t)
=
\strut\displaystyle
\frac{\epsilon_0\, a^3_0}{a^{3}(t)}
\\
\\
\epsilon^{\rm rad}(t)
=
\strut\displaystyle
\frac{\epsilon_0\, a^4_0}{a^{4}(t)}
\ .
\end{array}
\label{densityevolution}
\ee
The evolution of scale factors in cosmic time for expanding
($\uparrow$) and contracting ($\downarrow$) solutions are finally given by
\be
&&
\begin{array}{l}
a^{\rm rad}_{\uparrow\downarrow}(t)
=
\left(\gamma \pm 2\,\sqrt{M^{\rm rad}}\, t \right)^{1/2}
\\
\\
\strut\displaystyle
\frac{\d a^{\rm rad}_{\uparrow\downarrow}}{\d t}
=
\pm \frac{\sqrt {M^{\rm rad}}}{a^{\rm rad}(t)}
\end{array}
\label{arad}
\ee
and
\be
&&
\begin{array}{l}
a^{\rm dust}_{\uparrow\downarrow}(t)
=
\left(\delta\pm\frac{3}{2}\,\sqrt{M^{\rm dust}} \, t\right)^{2/3}
\\
\\
\strut\displaystyle
\frac{\d{a}^{\rm dust}_{\uparrow\downarrow}}{\d t}
=
\strut\displaystyle
\pm\sqrt{\frac{M^{\rm dust}}{a^{\rm dust}(t)}}
\ ,
\end{array}
\label{adust}
\ee
where, in the above r.h.s., the $+$ signs are for expansion and $-$
signs for contraction,
\be
\begin{array}{l}
M^{\rm rad}=\kappa\, a^4_0\,\epsilon^{\rm rad}_0
\\
\\
M^{\rm dust}=\kappa\, a^3_0\,\epsilon^{\rm dust}_0
\ ,
\end{array}
\ee
and $\gamma$ and $\delta$ integration constants
that determine the size of the scale factors at $t=0$.
Later, for convenience, we will set $\gamma = \delta = 1$ at $t=0$, so that
$\epsilon(0) = \epsilon_0$ and $a(0)=a_0=1$.
\par
Let us now consider the time-like shell $\Sigma$ at $r_\pm=r_\pm^{\rm s}(t_\pm)$
separating the two regions $\Omega_\pm$.
Clearly, metric continuity implies
\be
\rho=a_\pm(t_\pm)\,r_\pm^{\rm s}(t_\pm)
\ .
\ee
The inner and outer spaces are characterized by different physical parameters.
In particular, as one can see from Eq.~\eqref{B}, the shell's dynamics 
are determined by:
\begin{enumerate}
\item{The type of fluid inside the shell (its equation of state $w_-$);}
\item{The initial values $\ainz$ and $\einz$;}
\item{The type of fluid outside the shell (its equation of state $w_+$);}
\item{The initial values $\aoutz$ and $\eoutz$;}
\item{The shell surface density $\sigma$ (as a function of the radius $\rho$).}
\end{enumerate}
A given configuration of dust, radiation and surface density is admissible only if the
corresponding initial conditions are such that Eqs.~\eqref{drho0}, \eqref{dB0} and
\eqref{difference0} are satisfied.
\subsection{Time transformations and expansion}
The densities $\epsilon_\pm$ in Eq.~(\ref{densityevolution}) 
are given in terms of coordinate times $t_\pm$.
However, it is the time $\tau$ measured by an observer on the shell
which appears in the evolution equation~\eqref{evolution}.
Hence we need to find the transformation from $t_\pm$ to $\tau$.
Following Ref~\cite{bkt}, we recall that metric continuity~\eqref{cont}
implies~\footnote{There is a typo in Eq.~(B10) of Ref~\cite{bkt}:
$\rho^2$ is missing in the last term in the square root.}
\be
\!\!\!\!
\left. \frac{\d t_\pm}{\d \tau}\right|_{\Sigma}
\!\!\!
 =
\left\{ \frac{H\, \rho\, \dot{\rho}}{\Delta}
\left[
1
\pm
\sqrt{1+
\frac{\Delta^2-\Delta\left(\dot\rho^2+H^2\rho^2\right)}
{(H\, \rho\, \dot{\rho})^2}}
\right]
\right\}_{\pm}
\!\!\!\!\!
,
\label{tt}
\ee
in which $H$ is again the Hubble ``constant'', $\Delta=\kappa\,\epsilon\, \rho^2-1$,
and the expression within braces must be estimated on the two sides of the
shell~\footnote{Note the sign ambiguity $\pm$ in front of the square root just
reflects the double root of a second degree equation and is \emph{not\/}
associated with the interior or exterior regions.}.
Now, the above two equations should be solved along with Eq.~\eqref{evolution},
which makes it clear why it is impossible to obtain general analytic solutions.
\par
An important result can be obtained by considering the time when the bubble is
at rest, that is $t_\pm=t_{0\pm}$ and $\tau=\tau_0$, with $\dot\rho(\tau_0)=0$ and
$H_0 \equiv H(t_0)$, namely
\be
\left. \frac{\d t_\pm}{\d \tau}\right|_{\Sigma , 0}
=
\pm\sqrt{1-\frac{H^2_{0\pm}\,\rho^2_0}{\Delta_{0\pm}}}
\ .
\ee
From Eq.~\eqref{friedmann1} we see that $\Delta=H^2\rho^2-1$,
therefore
\be
\left. \frac{\d t_\pm}{\d \tau}\right|_{\Sigma , 0}
&\!\!=\!\!&
\frac{\pm 1}{\sqrt{{1-H^2_{0\pm}\,\rho^2_0}}}
=
\frac{\pm 1}{ \sqrt{{1-\kappa\,{\epsilon_0}_{\pm}\, \rho^2_0}}}
\nonumber
\\
&\!\!=\!\!&
\frac{\pm 1}{\sqrt{-{\Delta_{0\pm}}}}
\label{tdot}
\ ,
\ee
with the signs in the numerator simply reflecting the directions $t_\pm$ flow
relative to $\tau$.
It is clear that real solutions to Eq.~\eqref{tdot} exist only if 
\be
\Delta_{0\pm} < 0
\ .
\label{Delta0}
\ee
Remarkably, this is the same as stating that the energy density inside
the radius $\rho=\rho_0$ must not generate a black hole, as
one can easily check by considering the Schwarzschild radius
$r_{\rm S} = 2\,G_{\rm N}\,M$
with $M=(4\,\pi/3)\,\epsilon_0\, \rho_0^3$.
This is manifest when considering $\epsilon_{0-}$ inside the bubble,
but it must also hold for the energy density $\epsilon_{0+}$ outside the bubble.
For the outer region, this means the bubble must lie inside the Hubble radius,
$\rho_0<H_0^{-1}$.
Putting together these conditions tells us that the temporal coordinates are properly
transformed only within a causal region of the spacetime.
\par
In order to study how the bubble grows after nucleation, we can
expand $t=t(\tau)$ for short times about $t_{0\pm}$ and $\tau_0$.
Further, we want all times directed the same way, so we choose the $+$ sign
in the above expression and obtain, to linear order,
\be
t_{\pm}
\simeq
t_{0\pm}+\frac{\tau-\tau_0}{\sqrt{-{\Delta_{0\pm}}}}
\ ,
\label{transf}
\ee
where $t_{0\pm}$ are integration constants.
\par
Unfortunately, a first order expansion is not sufficient to study
the evolution of the bubble radius.
Since $\dot \rho_0= 0$, we need at least second order terms in $\tau$
to get significant results, which makes all expressions very cumbersome.
We shall therefore just consider a few specific cases, generalizing the exact
result~\eqref{vacuum} for a shell of constant surface energy in vacuum.
For such cases, our perturbative approach will yield exact
conditions for the bubble's existence, which we see as a clear advantage with
respect to purely numerical solutions.
Other advantages would be that having analytical expressions is a necessary
ingredient for quantum mechanical (or semiclassical) studies of these systems. 
Moreover, adapting our procedure to all possible combinations of fluids
in $\Omega_\pm$, and for more general shell surface density,
should be rather straightforward. 
\section{Radiation bubble inside dust}
\label{rad_dust}
The main idea in our approach stems from the observation that the (three)
fundamental (first order differential) equations~\eqref{evolution} and \eqref{tt}
contain six functions of the proper time $\tau$:
the shell radius $\rho$, its surface density $\sigma$, the two times $t_\pm$ and
the two Hubble functions $H_\pm$.
Once we choose the matter content inside $\Omega_\pm$ and on the shell, the Hubble
functions and surface density are uniquely fixed and we are left with the three unknowns
$\rho$ and $t_\pm$ (and a set of constraints for the initial conditions).
To determine these unknowns, we find it convenient to formally expand the shell radius
$\rho$ and Hubble functions $H_\pm$ for short (proper) time ``after the nucleation
of the bubble'' (when $\dot\rho_0=0$), and solve Eqs.~\eqref{evolution} and \eqref{tt}
order by order.
\par
Since expressions rapidly become involved, and a general treatment for
all combinations of matter content in $\Omega_\pm$ and shell surface density
would be hardly readable,
we shall only consider the specific case of a radiation bubble ($w_-=1/3$)
inside a region filled with dust ($w_+=0$).
We further assume the shell's surface density
\be
\sigma(\tau)=\sigma_0>0
\ee
is constant and positive.
Since one would expect the shell's density decreases as the shell's surface grows,
this assumption might appear rather strong.
However, it is the simplest way to ensure the junction remains of the ``black hole'' type,
and a more thorough discussion of this point can be found in Ref.~\cite{bkt}.
In order to keep the presentation uncluttered, we also set $\kappa=1$ from now on
and regard all quantities as dimensionless (tantamount to assuming they are
rescaled by suitable powers of $\kappa=\lp/\mpl$).
This means that densities will be measured in Planck units,
that is $\epsilon=1$ corresponds to the Planck density
$\epsilon_{\rm P}=\mpl/\lp^3=\lp^{-2}$ and $\sigma=1$ to
$\sigma_{\rm P}=\mpl/\lp^2=\lp^{-1}$.
Likewise, $\rho=1$ is the Planck length $\lp$.
We also express the shell surface density and radiation energy density
as fractions of $\ezd>0$,
\be
\ezr=\ezd\,x
\ ,
\qquad
\sigma_0=\sqrt{\ezd}\,y
\ ,
\label{xy}
\ee
with $0\le x\le 1$ and $y\ge 0$.
\par
It is natural to choose $\tau_0=0$, and then proceed to analyze
Eqs.~\eqref{evolution} and \eqref{tt} by formally expanding all relevant time-dependent
functions in powers of $\tau-\tau_0=\tau$:
\par
{\bf Step 1)}
since $\dot \rho_0 = 0$ [see Eq.~\eqref{drho0}], we can formally write the bubble radius as
\be
\rho(\tau)
=
\rho_0 + \frac{1}{2}\,\ddot\rho_0 \,\tau^2
+\mathcal{O}(\tau^3)
\ ,
\label{rhos}
\ee
where $\rho_0$ and $\ddot\rho_0$ are parameters to be determined.
In particular, from Eqs.~\eqref{B} and \eqref{rho0}, we obtain the (not yet final)
expression
\be
\rho_0
&\!\!=\!\!&
\frac{3\,\sigma_0}{\sqrt{\left(\eoutz+\einz+9\,\kappa\,\sigma_0^2/4\right)^2-4\,\einz\, \eoutz}}
\nonumber
\\
&\!\!=\!\!&
\frac{3\,(\ezd)^{-1/2}\,y}
{\sqrt{\left(1+x+9\,y^2/4\right)^2-4\,x}}
\ ,
\label{rho00}
\ee
which only depends on $\epsilon_{0\pm}$ and $\sigma_0$.
More precisely, $\ezd$ sets the overall scale of the shell radius
and the fractions $x$ and $y$ defined in Eq.~\eqref{xy} the detailed form.
\par 
We next obtain $t_\pm=t_\pm(\tau)$ by solving Eq.~(\ref{tt}).
However, for this purpose we need the Hubble parameters $H_\pm$ as functions of $\tau$,
whereas they explicitly depend on $t_\pm$:
\par
{\bf Step 2)}
we replace $H$ in Eq.~(\ref{tt}) with the formal expansion
\be
\h
=
\h_0 + \dot{\h}_0\, \tau
+\mathcal{O}(\tau^2)
\ ,
\label{HH}
\ee
where $\h_0$ and $\dot{\h}_0$ are unknown constant quantities
to be determined by consistency.
By expanding the right hand side of Eq.~(\ref{tt}) to first order
in $\tau$ and then integrating, we obtain $t$ to second order in $\tau$,
\be
\!\!
 t_\pm
 \simeq
 t_{0\pm}
 +
 \frac{\tau}{\sqrt{-\Delta_{0 \pm}}}
 +
 \frac{\rho_0\,\h_{0\pm}}{2\,\Delta_{0\pm}} 
\left(
{\ddot \rho}_0-\frac{\rho_0\,{\dot \h}_{0\pm} }{\sqrt{-\Delta_{0\pm}}}\right)
\tau^2
,
\label{secondorder}
\ee
where $\rho_0$ must now be understood as the expression given in Eq.~\eqref{rho00}
and $t_{0\pm}$ are integration constants we can set to zero without loss
of generality.
In fact, let us assume we are at rest in an expanding (or contracting) universe,
corresponding to the old exterior phase (with parameters $\epsilon_{0+}$ and $H_{0+}$),
and measure a time $t_{+}'$ from the ``Big Bang'' (or beginning of collapse)
of this exterior universe [$a_+(t_+'=0)=0$ or $a_+(t_+'=0)>a(t_+')$ for $t'_+>0$,
respectively].
If, for instance, a new phase bubble arises at rest at the instant $t_+'=t_{0+}'$,
we can define $t_{+}=t_+'-t'_{0+}$, so that the bubble is created at $t_+=0$,
and also set $t_{0-}=0$, since an ``inner time'' is meaningless before
any ``inner part'' exists.
For different pictures, similar arguments can likewise be formulated.
\par
{\bf Step 3)}
From Eq.~(\ref{arad}) and (\ref{adust}), we choose an expanding radiation interior
and contracting or expanding dust exterior,
\be
a_-=
a^{\rm rad}_\uparrow(t_-)
\ ,
\qquad
a_+=a^{\rm dust}_{\downarrow\uparrow}(t_+)
\ ,
\ee
set $a_{0\pm}=1$ and
express $t_\pm$ according to Eq.~(\ref{secondorder}).
In so doing, $a_\pm$ and $\d a_\pm/\d t_\pm$ become explicit functions of
$\tau$ containing $\rho_0$, $\h_{0\pm}$ and $\dot{\h}_{0\pm}$.
For consistency with Eq.~(\ref{HH}), we must therefore require
\be
H(\tau)=
\frac{1}{a}\,
\frac{\d a}{\d t}
=\h(\tau)
\ ,
\label{eqH}
\ee
with $H_-=\h_->0$ and $H_+=\h_+<0$ for collapsing dust, or $H_+=\h_+>0$ for
expanding dust.
\par
{\bf Step 4)}
To zero order in $\tau$, Eq.~\eqref{eqH} gives rise to a first-order equation for $\h_0$,
\be
H_0
=
\frac{1}{a_0}
\left.
\frac{\d a}{\d t}
\right|_{\tau=0}
=
\left.
\frac{\d a}{\d t}
\right|_{\tau=0}
= \h_0
\ .
\label{eqH0}
\ee
The solutions are uniquely given by
\be
\h_0^{\uparrow\downarrow}
=
\pm\sqrt{\epsilon_0}
\ ,
\label{h0}
\ee
in which the $\uparrow$ and $+$ sign (respectively $\downarrow$ and $-$ sign) refer
to expanding (contracting) solutions, i.e.~solutions with increasing (decreasing)
scale factor, as before~\footnote{Note that, for example, the Hubble parameter for the expanding
interior phase will carry a second subscript sign and will then be denoted as $\h_-^\uparrow$,
where the subscript $-$ indicates the inner region and the apex $\uparrow$ stands for expansion.}.
Note this result also follows directly from the Friedmann equation~\eqref{friedmann1}
for $\tau=t=0$.
\par
To first order in $\tau$, one analogously obtains
\be
\dot\h_0^{\uparrow\downarrow}
=
-\frac{n\,\epsilon_0}{\sqrt{1-\rho_0^2\,\epsilon_0}}
\ ,
\ee
with $n=2$ for radiation and $n=3/2$ for dust, and $\rho_0$ must again
be understood as the expression given in Eq.~\eqref{rho00}.
\par
{\bf Step 5)}
Replace $\rho_0$ from Eq.~\eqref{rho00} and the chosen combination of 
Hubble parameters~\eqref{h0} inside $\dot B_0$, which must then satisfy
Eq.~\eqref{dB0}.
This equation will only contain $\einz=\ezr$, $\eoutz=\ezd$ and $\sigma_0$,
so that it can be used to determine
$\sigma_0=\sigma_0(\einz,\eoutz)$.
In particular, introducing the fractions in Eq.~\eqref{xy}, we obtain
\be
\dot B_0=\ezd\,\dot b_0(x,y)
\ ,
\ee
from which it appears that the dust energy density just sets the overall scale like
in Eq.~\eqref{rho00}.
For any given values of $\ezd$, the shell surface density is instead determined by
the radiation energy density according to
\be
\dot b_0(x,y)=0
\ ,
\label{rhs0}
\ee
which, for the cases of interest, is a fourth-order algebraic equation for $y$.
Analytic solutions can be found (in suitable ranges of $x$),
which we denote as $\bar y=\bar y(x)$, so that the allowed surface densities are
given by
\be
\bar \sigma_0
=\sqrt{\ezd}\,\bar y
\ .
\ee
\par
{\bf Step 6)}
Replace the above surface density $\bar\sigma_0$ into the initial radius~\eqref{rho00}
and obtain its final expression,
\be
\bar \rho_0
=
\frac{3\,(\ezd)^{-1/2}\,\bar y}
{\sqrt{\left(1+x+9\,\bar y^2/4\right)^2-4\,x}}
\ ,
\ee
which can then be used to determine the final forms of $\dot\h_{0\pm}$ and the scale
factors $a_\pm$ to first order in $\tau$.
\par
{\bf Step 7)}
One must now check that $\bar\sigma_0$ and $\bar\rho_0$ satisfy all of the initial
constraints and lead to valid time transformations~\eqref{tt},
at least for some values of $\ezr$ and $\ezd$.
If not all of these conditions are met, one must conclude the corresponding physical system
may not exist.  
Moreover, we note the condition~\eqref{Delta0} requires $\epsilon_0\lesssim \epsilon_{\rm P}$
in order to have a (semi)classical bubble with $\rho_0\gtrsim\lp$.
In the following Section, one should therefore consider only dust and radiation energy densities
$\epsilon_0\ll\epsilon_{\rm P}$ and look at the limiting case $\epsilon_0\simeq\epsilon_{\rm P}$
as a glimpse into the quantum gravity regime.
\par
If a consistent solution for $\bar\sigma_0$ and $\bar\rho_0$ exists,
one can proceed to determine higher orders terms (in $\tau$).
However, due to the increasing degree of complexity of the resulting expressions,
we shall not go any further here.
We instead present our findings for the two cases of interest separately.
\subsection{Collapsing dust}
This case is defined by choosing the scale factors
\be
a_-=
a^{\rm rad}_\uparrow(t_-)
\ ,
\qquad
a_+=a^{\rm dust}_{\downarrow}(t_+)
\ ,
\ee
and proceeding as described above.
We can then prove that this case does not admit solutions, in general,
by simply analyzing the constraint~\eqref{rhs0},
\be
\dot b_0
&\!\!\propto\!\!&
16\,x^2\!
\left(3+4\,x^{3/2}\right)
+
4\,x^{3/2}\!
\left(4-9\,y^2\right)
\!\!
\left(4-8\,x-9\,y^2\right)
\nonumber
\\
&&
+3\left(4+9\,y^2\right)\!
\left(4-8\,x+9\,y^2\right)
=0
\ ,
\ee  
which admits the four solutions
\be
\bar y_{\pm\pm}
=
\pm\frac{2}{3}\,
\sqrt{(1-x)\,\frac{2\,x^{3/4}\pm i\,\sqrt{3}}{2\,x^{3/4}\mp i\,\sqrt{3}}}
\ .
\ee
However, for $x\not= 1$, all the $\bar y_{\pm\pm}$ are complex
and a complex surface density is obviously unphysical.
One is then apparently left with the only trivial case $x=1$,
corresponding to $\bar y=0$ and 
\be
\bar\rho_0
=
(\ezd)^{-1/2}\sqrt{\frac{3+4\,x^{3/2}}{3+4\,x^{1/2}}}
=
1
\ .
\ee
This case however does not satisfy all the required constraints.
For example, Eq.~\eqref{difference0} for $\sigma_0=0$ yields
\be
\ezd>\ezr=x\,\ezd
\ ,
\ee
which clearly contradicts $x=1$.
Correspondingly, the time transformations~\eqref{tt} are not well-defined,
because $\dot {\bar t}_{0+}=\dot{\bar t}_+(\tau=0;x)$ is complex for $0<x<1$
and both $\dot t_{0\pm}$ diverge for $x\to 1$. 
\par
The overall conclusion is thus that expanding radiation bubbles 
with a turning point of minimum radius  cannot be matched with a
collapsing dust exterior.
\subsection{Expanding dust}
\begin{figure}[t]
\centering
\includegraphics[width=7cm]{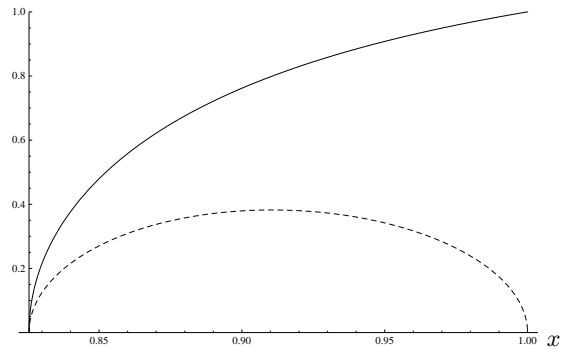}
{$x$}
\caption{Plot of $\bar \sigma_0/\sqrt{\ezd}=\bar y_{+-}(x)$
(magnified by a factor of $10$ for convenience,
dashed line) and corresponding $\bar\rho_0/(\ezd)^{-1/2}$ (solid line)
in the range~\eqref{range}.
}
\label{exp}
\end{figure}
\begin{figure}
\centering
\includegraphics[width=7cm]{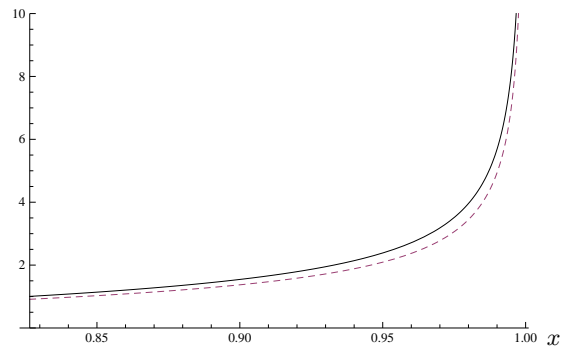}
{$x$}
\caption{Plot of $\dot{\bar t}_{0+}$ (solid line) and $\dot{\bar t}_{0-}$
(dashed line) for $y=\bar y_{+-}(x)$ in the range~\eqref{range}.
}
\label{expH}
\end{figure}
\begin{figure}
\centering
\includegraphics[width=7cm]{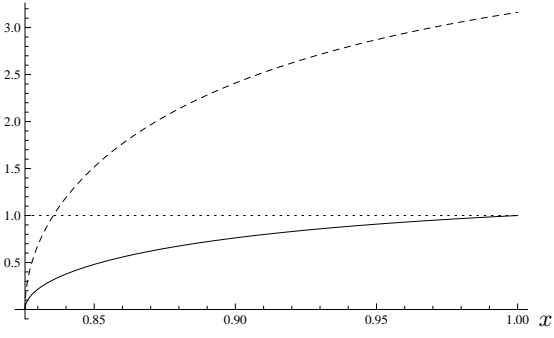}
{$x$}
\caption{Plot of $\bar\rho_0$ for $y=\bar y_{+-}(x)$ with $\ezd=\epsilon_{\rm P}/10$
(dashed line) and $\ezd=\epsilon_{\rm P}$ (solid line)
in the range~\eqref{range}.
Only values above $\lp=1$ represent acceptable semiclassical radii.}
\label{rhoP}
\end{figure}
\begin{figure}
\centering
\includegraphics[width=7cm]{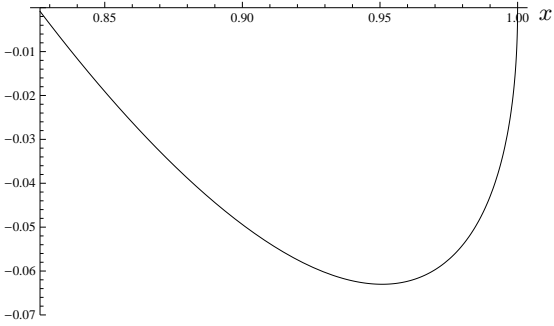}
\raisebox{4cm}{$x$}
\caption{Plot of $\bar C_0/\bar M_0^{\rm dust}$ for $\bar\sigma_0>0$ and $y=\bar y_{+-}(x)$
in the range~\eqref{range}.
}
\label{C0p}
\end{figure}
This case is defined by choosing the scale factors
\be
a_-=
a^{\rm rad}_\uparrow(t_-)
\ ,
\qquad
a_+=a^{\rm dust}_{\uparrow}(t_+)
\ .
\ee
The crucial task is again to solve the constraint in Eq.~\eqref{rhs0},
namely
\be
\dot b_0
&\!\!\propto\!\!&
16\,x^2\!
\left(3-4\,x^{3/2}\right)
-
4\,x^{3/2}\!
\left(4-9\,y^2\right)
\!\!
\left(4-8\,x-9\,y^2\right)
\nonumber
\\
&&
-3\left(4+9\,y^2\right)\!
\left(4-8\,x+9\,y^2\right)
=0
\ ,
\ee  
admitting the four solutions
\be
\bar y_{\pm\pm}
=
\pm\frac{2}{3}\,
\sqrt{(1-x)\,\frac{2\,x^{3/4}\pm \sqrt{3}}{2\,x^{3/4}\mp \sqrt{3}}}
\ ,
\ee
which are real for 
\be
x_{\rm min}
=
\left(\frac{3}{4}\right)^{2/3}
< x< 1
\ .
\label{range}
\ee
We discard the negative solutions $\bar y_{-\pm}$ associated to negative
surface densities and just analyze the positive cases $\bar y_{+\pm}$.
It is then easy to see that $\bar y_{++}$ leads to a surface density that diverges
for $x\to x_{\rm min}$, and is always too large to satisfy the condition~\eqref{difference0},
since
\be
1-x-\frac{9}{4}\,\bar y_{++}^2
=
\frac{\sqrt{3}\left(x-1\right)}{2\,x^{3/4}-\sqrt{3}}
<0
\ ,
\ee
in the range~\eqref{range}.
In the limit for $x\to 1$, $\bar y_{++}\to 0$, however, Eq.~\eqref{difference0} is still violated
in the strict sense and one can in fact show that $\dot {\bar t}_{0+}$ diverges.
\par
The only solution which appears consistent is therefore
\be
\bar\sigma_0
&\!\!=\!\!&
\sqrt{\ezd}\,\bar y_{+-}
\nonumber
\\
&\!\!=\!\!&
\frac{2}{3}\,\sqrt{\ezd}\,
\sqrt{(1-x)\,\frac{2\,x^{3/4}- \sqrt{3}}{2\,x^{3/4}+ \sqrt{3}}}
\ ,
\label{s0fin}
\ee
with $x$ again in the range~\eqref{range}. 
This expression yields a vanishing surface density for the limiting values
$x\to 1$ and $x\to x_{\rm min}$ (see Fig.~\ref{exp}) and further
satisfies the condition~\eqref{difference0},
\be
1-x-\frac{9}{4}\,\bar y_{+-}^2
=
\frac{\sqrt{3}\left(1-x\right)}{2\,x^{3/4}+\sqrt{3}}
>
0
\ .
\ee
The corresponding initial bubble radius is an increasing
function of $x$ (see Fig.\ref{exp}),
\be
\bar\rho_0
=
(\ezd)^{-1/2}
\sqrt{\frac{4\,x^{3/2}-3}{x\left(4\,x^{1/2}-3\right)}}
<
(\ezd)^{-1/2}
\ ,
\label{r0fin}
\ee
with $\bar\rho_0(x\to 1)=(\ezd)^{-1/2}$.
Further, the products
\be
\ezd\,\bar\rho_0^2<1
\quad
{\rm and}
\quad
\ezr\,\bar\rho_0^2<1
\ ,
\ee
for $x_{\rm min}<x<1$, as required by the condition~\eqref{Delta0}.
In fact the initial time derivatives $\dot{\bar t}_{0\pm}$ are well defined in this
range (see Fig.~\ref{expH}) and only diverge for $x\to 1$.
Note the above initial radius can be larger than $\lp$ only if
$\ezd < \epsilon_{\rm P}$
and for sufficiently large $x$, since $\bar\rho_0\to 0$ for $x\to x_{\rm min}$
(see, for example, Fig.~\ref{rhoP}).
\par
Finally, let us check if one can use the process of bubble nucleation
to describe a phase transition from dust to radiation for the matter
inside the sphere of radius $\bar\rho_0$, accompanied by the creation
of a layer of non-vanishing surface density $\bar\sigma_0$.
From Eqs.~\eqref{s0fin} and \eqref{r0fin}, one has
\be
\bar C_0
\equiv
\bar M_0^{\rm dust}
-
\bar M_0^{\rm rad}
-
\bar E_0^{\Sigma}
<0
\ ,
\label{C0}
\ee
which means the dust energy inside the sphere of radius $\bar\rho_0$
at time of bubble formation,
$\bar M_0^{\rm dust}=(4\,\pi/3)\,\bar\rho_0^3\,\ezd$, is not sufficient
to produce the radiation energy $\bar M_0^{\rm rad}=(4\,\pi/3)\,\bar\rho_0^3\,\ezr$
and surface energy $\bar E_0^{\Sigma}=4\,\pi\,\bar\sigma_0\,\rho_0^2$.
An extra source is thus needed to provide the energy $-\bar C_0>0$.
The reverse process of collapsing radiation reaching a minimum size
$\rho=\bar\rho_0$ and then turning into collapsing dust would instead be
energetically favored, with the amount of energy $-\bar C_0$ now being released.
Of course, in order to support this kind of argument, the extra contribution
should be a small perturbation on the given background,
\be
{|\bar C_0|}\ll{\bar M_0^{\rm dust}}
\ ,
\ee
since it was not included in the dynamical equations.
From Fig.~\eqref{C0p}, we expect this is indeed a very good approximation
since $0<-\bar C_0\lesssim0.06\,\bar M_0^{\rm dust}$.
\section{Conclusions and outlook}
\label{conc}
In this paper, we have analyzed bubbles of radiation whose time-like surface starts
to expand inside collapsing or expanding dust with vanishing initial rate, and with the further
(simplifying) assumptions that the bubble's surface density is constant and positive,
and both interior and exterior are spatially flat.
These bubbles generalize the simplest self-gravitating case of a shell with
constant surface density expanding in vacuum, for which the exact trajectories
are known~\cite{israel,ansoldi}.
These generalizations are of potential interest both for the physics of the
early universe and the description of astrophysical processes.
However, although the general formalism was already developed a long time
ago~\cite{israel}, and the dynamics are ruled by apparently simple equations~\cite{bkt},
finding explicit solutions is not straightforward.
\par
By developing an approach to obtain analytical expressions for the evolution of the
bubble radius in the shell's proper time, $\rho=\rho(\tau)$ with $\dot\rho(\tau=0)=0$,
we determined the conditions which allow for the existence of such configurations.
Although our approach is perturbative (with an expansion for short times after nucleation), 
the conditions for the bubble's existence are exact, which is a clear advantage with
respect to purely numerical solutions.
We then found that expanding radiation bubbles of constant
surface density may not be matched to a collapsing dust exterior.
More precisely, we found that inside collapsing dust there may not be a bubble of
radiation whose surface ever reaches vanishing speed of expansion at finite radius.
Bubbles whose radius admits a turning point are instead allowed inside an expanding
dust-dominated universe.
They further can be used to model a phase transition from radiation
to dust if an external source provides part of the the energy required to build
the shell, or the converse process with release of energy (albeit, of an amount small
enough to leave the background configuration unaffected).
\par
Let us clarify this point about energy conservation.
The fundamental Eqs.~\eqref{evolution} and \eqref{tt} are just a different form of the
junction equations~\eqref{cont} and \eqref{K} which, in turn, follow from the
Einstein equations.
Conservation of the energy-momentum in a given system is therefore guaranteed.
However, when we use bubbles to model a phase transition, we are considering
the possibility that a region of space filled with dust be replaced by radiation
enclosed inside an expanding shell, or the reverse process.
Technically, we are therefore considering two different systems:
one with dust and one with a bubble of radiation within a shell of positive
surface density whose radius evolves along a trajectory with a turning point
(zero speed at finite minimum radius).
The total energy in the two configurations differ by the amount $\bar C_0$
defined in Eq.~\eqref{C0}, and a (quantum) transition between them
would therefore violate energy conservation and be suppressed in the
semiclassical regime.
By looking at Fig.~\ref{C0p}, we however see that $|\bar C_0|\ll \bar M_0^{\rm dust}$.
One may thus argue the unspecified matter contribution carrying the energy $|\bar C_0|$
should be well approximated as a perturbation with respect to the dust and radiation,
with its backreaction on the chosen configuration consistently negligible.
If so, one can further speculate if the extra energy required to nucleate radiation
could be provided by pressure in the initial cloud or by the decay of a region
of false vacuum (with vacuum energy or cosmological constant $\Lambda_+$)
to true vacuum (with cosmological constant $\Lambda_-<\Lambda_+$), like in the
seminal Refs.~\cite{coleman}. 
\par
The fact that no consistent solution was found inside collapsing dust does not
mean that expanding radiation bubbles may not be produced at all in this context,
which includes, for example, the collapsing core of a
supernova or other astrophysical processes leading to black hole formation.
In fact, the situation might change if one, for instance (and more realistically),
includes matter pressure or a radius-dependent surface density,
$\sigma=\sigma(\rho)$.
This observation thus brings us to briefly comment on the possible generalizations
and extensions of the present work, which include the just mentioned
non-constant $\sigma$, as well as different combinations of matter inside
and outside the shell, and cosmological constant(s) $\Lambda_\pm$.
Moreover, one might like to consider the vacuum inside the shell and radiation outside
(with or without $\Lambda_\pm$) and apply the corresponding results to
the thick shell model previously studied in Refs.~\cite{ms}.
\par
Finally, our analysis is entirely based on classical General Relativity and
no attempt was made to compute the quantum mechanical ``tunneling'' probability
for radiation bubbles to come into existence (or convert to dust).
Such an analysis requires the (effective Euclidean) action to be integrated along the
(classically forbidden) trajectory for the bubble radius $\rho$ to go from $0$ to
$\rho_0$~\cite{farhi}, whose construction is clearly no easy task,
given the classical trajectories are so difficult to determine.
Nonetheless, another advantage of our approach is that it provides analytical
(albeit perturbative) expressions, which is a property one needs for
any quantum mechanical (or semiclassical) studies of these systems. 
Of course, energy densities above the Planck scale would not be meaningful
in this context, since one then has no guarantee the dynamical equations derived
from General Relativity can be trusted at all. 
\acknowledgments
We would like to thank G.L.~Alberghi and S.~Ansoldi for many discussions about the topic.
R.C.~and A.O.~are supported by INFN grant BO11. 
\end{document}